\documentclass{article}

\usepackage{microtype}
\usepackage{graphicx}
\usepackage{subfigure}
\usepackage{booktabs} 

\usepackage{hyperref}

\usepackage[accepted]{ml4astro2025}
\usepackage{amsmath}
\usepackage{amssymb}
\usepackage{mathtools}
\usepackage{amsthm}

\usepackage[capitalize,noabbrev]{cleveref}

\usepackage{multirow} 

\theoremstyle{plain}

\theoremstyle{definition}

\theoremstyle{remark}

\usepackage[textsize=tiny]{todonotes}

\mlforastrotitlerunning{Astro-MoE: Mixture of Experts for Multiband Astronomical Time Series}

\begin{document}

\twocolumn[
\mlforastrotitle{Astro-MoE: Mixture of Experts for Multiband Astronomical Time Series}




\begin{mlforastroauthorlist}
\mlforastroauthor{Martina Cádiz-Leyton}{edin}
\mlforastroauthor{Guillermo Cabrera-Vives}{udec,uds,ger,mas}
\mlforastroauthor{Pavlos Protopapas}{hv}
\mlforastroauthor{Daniel Moreno-Cartagena}{udec}
\mlforastroauthor{Ignacio Becker}{hv}

\mlforastroaffiliation{edin}{Edinburgh Futures Institute, University of Edinburgh, UK}
\mlforastroaffiliation{udec}{Department of Computer Science, Universidad de Concepción, Edmundo Larenas 219, Concepción, Chile}
\mlforastroaffiliation{uds}{Center for Data and Artificial Intelligence, Universidad de Concepción, Edmundo Larenas 310, Concepción, Chile}
\mlforastroaffiliation{ger}{Heidelberg Institute for Theoretical Studies, Heidelberg, Baden-Württemberg, Germany}
\mlforastroaffiliation{mas}{Millennium Institute of Astrophysics (MAS), Nuncio Monseñor Sotero Sanz 100, Of. 104, Providencia, Santiago, Chile}
\mlforastroaffiliation{hv}{John A. Paulson School of Engineering and Applied Sciences, Harvard University, Boston, MA, USA}

\mlforastrocorrespondingauthor{Martina Cádiz-Leyton}{m.a.cadiz@sms.ed.ac.uk}
\mlforastrocorrespondingauthor{Guillermo Cabrera-Vives}{guillecabrera@inf.udec.cl}
\mlforastrokeywords{Machine Learning, ICML}
\vskip 0.3in
\end{mlforastroauthorlist}
] 



\printAffiliationsAndNotice{}  

\begin{abstract}
Multiband astronomical time series exhibit heterogeneous variability patterns, sampling cadences, and signal characteristics across bands. Standard transformers apply shared parameters to all bands, potentially limiting their ability to model this rich structure. In this work, we introduce Astro-MoE, a foundational transformer architecture that enables dynamic processing via a Mixture of Experts module. We validate our model on both simulated (ELAsTiCC-1) and real-world datasets (Pan-STARRS1).

\end{abstract}

\section{Introduction}
\label{sec:Introduction}
After a decade of breakthroughs enabled by single-epoch surveys, astronomy is transitioning toward a new era defined by multi-epoch observations. This shift has fueled the rapid growth of time-domain astronomy, which focuses on studying celestial objects and phenomena whose properties evolve over time through wide-field surveys that repeatedly image large areas of the sky \cite{graham2012data, kasliwal2019growth}. The upcoming Legacy Survey of Space and Time (LSST; \citealt{ivezic2019lsst}), beginning full operations by the end of 2025, will accelerate this trend by generating multiband time series data for approximately 20 billion sources with unprecedented depth, cadence, and volume.

The intrinsic complexity of time-domain data (characterized by heterogeneous sampling, irregular cadences, and inter-band dependencies) has motivated significant advances in representation learning for astronomical time series (e.g. \citealt{protopapas2017recurrent, charnock2017deep, naul2018recurrent, park2021inferring, donoso2022astromer, pan2024astroconformer, becker2025multiband}). Transformer architectures \cite{vaswani2017attention} have shown particular promise for modeling such irregular sequential data, achieving state-of-the-art results within astronomy across tasks such as denoising \cite{morvan2022don}, classification \cite{pimentel2022deep, allam2024paying}, regression \cite{zhang2024spt}, and uncertainty estimation \cite{cadiz2024uncertainty}, even in data-scarce scenarios \cite{moreno2023positional, donoso2025astromer}. A core challenge in this domain is learning unified embeddings that compactly encode both temporal evolution and spectral characteristics across photometric bands. Such representations not only enhance performance on fundamental astronomical tasks (e.g., variable star classification, redshift regression) but also enable effective integration into multimodal frameworks \cite{rizhko2024self, lanusse2023astroclip, parker2024AstroCLIP}. For instance, recent models such as ATAT \cite{cabrera2024atat} demonstrate how combining light curve embeddings with auxiliary metadata can accelerate scientific discovery. 

However, conventional transformer-based architectures face fundamental limitations when processing multiband time series. Transformers apply homogeneous processing to all inputs (i.e., reuse the same parameters for all inputs), which is suboptimal for astronomical data where each photometric band probes distinct astrophysical processes, exhibits unique sampling characteristics, and manifests independent variability behaviors. This diversity suggests that uniform weight-sharing mechanisms may not adequately capture band-specific features and complex inter-band interactions. Mixture of Experts (MoE) architectures offer a compelling solution through dynamic, input-dependent routing to specialized subnetworks \cite{masoudnia2014mixture, shazeer2017outrageously}. MoE-based models select different parameters for each example, resulting in sparsely-activated models with more parameters but constant computational cost. This capability is ideally suited to multiband astronomical data, where experts can learn band-specific representations while maintaining shared temporal knowledge. While recent works like Time-MoE \cite{shiscaling} and Moirai-MoE \cite{liumixture} highlight the potential of expert routing in temporal data, their adaptation to the specific challenges of irregular, multiband astronomical time series remains an area of ongoing interest, with opportunities for further methodological development.

In this work, we introduce Astro-MoE, a pretrained transformer that incorporates sparsely-gated MoE modules in the architecture. This design enables the processing of multiband time series, producing embeddings that are both robust and informative for downstream tasks. Our architecture addresses the challenges of astronomical time series by allowing different experts to specialize in different variability patterns while maintaining the ability to model complex inter-band correlations.



\section{Methods}
\label{sec:Methods}


The Astro-MoE model extends the Astromer framework~\cite{donoso2022astromer}, a self-supervised light curve transformer originally designed for single-band data, to a multiband setting. Each astronomical object is represented as a sequence of observations across multiple photometric bands. For each band \( b \) and time step \( j \), the input includes a flux measurement \( \mu_{j,b} \) and its associated uncertainty \( \sigma_{j,b} \). These values are combined into an input vector \( x_{j,b} = (\mu_{j,b}, \sigma_{j,b}) \), which represents the brightness and its error at time \( j \) in band \( b \). To construct the model input, each band is encoded as a fixed-length sequence of \( x_{j,b} \) vectors, with zero-padding applied as needed. The resulting sequences are then concatenated across bands to form a unified representation \( x \), which preserves temporal ordering and captures band-specific information.

\subsection{Mixture of Experts}

\begin{figure}[t]
\begin{center}
\includegraphics[width=8cm]{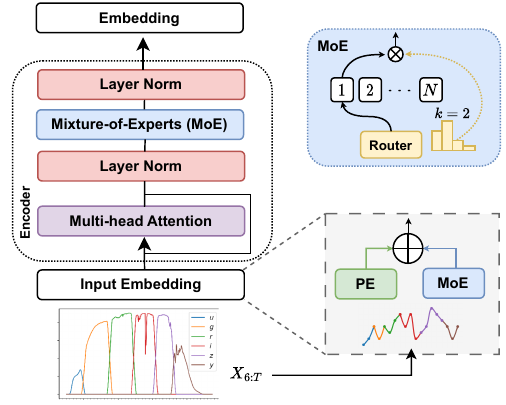}
\caption{Astro-MoE architecture diagram with the traditional positional encoding.}
\label{fig:astromer_moe}
\end{center}
\vskip -0.10in
\end{figure}

As illustrated in Figure~\ref{fig:astromer_moe}, our model adopts an encoder-only transformer architecture enhanced with sparse Mixture-of-Experts (MoE) layers~\citep{shazeer2017outrageously,mu2025comprehensive}. The MoE module is integrated into two key components: (1) the input embedding stage, and (2) the attention blocks, where it replaces the standard feedforward network (FFN) sublayer. Each MoE layer contains \( N_{\mathrm{experts}} \) parallel experts, with sparse routing controlled by a learnable gating function. Here, \( N_{\mathrm{experts}} \) denotes the number of experts in the layer, which can differ between components.

Given an input vector \( x \in \mathbb{R}^{d_{\mathrm{in}}} \), where \( d_{\mathrm{in}} \) is the dimensionality of the input (e.g., 2 for brightness and uncertainty pairs), the gating network computes a score for each expert:
\begin{equation}
g(x) = W_g x + b_g, \quad g: \mathbb{R}^{d_{\mathrm{in}}} \rightarrow \mathbb{R}^{N_{\mathrm{experts}}},
\end{equation}
where \( W_g \) and \( b_g \) are the learnable weights and bias of the gating network. To induce sparsity, only the top-$k$ scoring experts are selected using a $\mathrm{TopK}$ operator, and a softmax is applied to compute normalized selection weights:
\begin{equation}
G(x) = \mathrm{softmax}(\mathrm{TopK}(g(x), k)),
\end{equation}
where \( k \) is the number of experts selected per input. The operator \(\mathrm{TopK}(v, k)_i\) retains the top-$k$ values in \(v\) and masks the rest with \(-\infty\):
\[
\mathrm{TopK}(v, k)_i = 
\begin{cases}
v_i, & \text{if } v_i \text{ is among the top-}k \text{ elements}, \\
-\infty, & \text{otherwise}.
\end{cases}
\]
This yields a sparse selection vector $G(x) \in \mathbb{R}^{N_{\mathrm{experts}}}$ with nonzero weights only for the selected experts (i.e., $G(x)_e \neq 0$ if and only if expert $e$ is selected). The final MoE output is then computed as:
\begin{equation}
\mathrm{MoE}(x) = \sum_{e \in \mathrm{Top\text{-}k}} G(x)_e \cdot E^{(e)}(x),
\end{equation}
where $E^{(e)}(x)$ is the output of expert $e$.

In the input embedding stage, each expert \( E^{(e)} \) is implemented as a linear transformation:
\begin{equation}
E^{(e)}(x) = W^{(e)} x, \quad W^{(e)} \in \mathbb{R}^{d_{\mathrm{model}} \times d_{\mathrm{in}}}.
\end{equation}
where \( d_{\mathrm{model}} \) is the internal model dimension used by the transformer. Unlike prior approaches that apply a uniform transformation across all bands~\citep{donoso2022astromer,cabrera2024atat}, we project the brightness-uncertainty pairs \( (x_{j,b}, \sigma_{j,b}) \in \mathbb{R}^2 \) into \( d_{\mathrm{model}} \) using a MoE module. We set \( N_{\mathrm{experts}} = 6 \) (one per photometric band) and \( k = 2 \), matching the six-band pretraining setup. This configuration promotes expert specialization across different observational regimes.

Within the attention blocks, each expert is implemented as a two-layer FFN, replacing the standard FFN sublayer. This choice is inspired by prior large-scale MoE transformer models~\citep{lepikhin2020gshard,fedus2022switch}, which demonstrate that FFN layers often exhibit sparse and task-specific activation patterns, making them well-suited for expert-based specialization. In this setting, we use \( N_{\mathrm{experts}} = 8 \) and select the top-$k = 2$ experts per token.

To encourage balanced expert utilization and avoid expert collapse, we incorporate a simplified version of the load balancing loss from \citet{shazeer2017outrageously}:
\begin{equation}
\mathcal{L}_{\text{aux}} = N_{\text{experts}} \cdot \sum_{e=1}^{N_{\text{experts}}} \bar{p}_e \cdot \bar{f}_e,
\end{equation}
where \( \bar{p}_e \) is the mean routing probability assigned to expert \( e \), and \( \bar{f}_e \) is the empirical fraction of tokens routed to that expert. This auxiliary loss is computed independently for each MoE layer and summed across all such layers. The total auxiliary loss is then scaled by a fixed coefficient \( \lambda = 0.01 \) and added to the main task loss (e.g., classification or regression). The scaling factor ensures that the auxiliary term promotes expert diversity without overpowering the primary learning objective.

\subsection{Positional Embedding}
Encoding temporal information is a crucial component in modeling light curves, as observations are irregularly spaced in time. Recent work has shown that the choice of positional encoding (PE) can significantly affect the performance of transformer-based models for time series~\citep{moreno2023positional}. Therefore, we explore two temporal encoding strategies, both integrated at the input embedding stage, before the self-attention layers.

The first approach follows the original transformer formulation~\citep{vaswani2017attention}, in which observation times are encoded using fixed sine and cosine functions at varying frequencies:
\begin{equation}
\mathrm{PE}_{i}(t_{j,b}) =
\begin{cases}
\sin(t_{j,b} \cdot \omega_i), & i \text{ even}, \\
\cos(t_{j,b} \cdot \omega_i), & i \text{ odd},
\end{cases}
\quad \omega_i = \frac{1}{1000^{2i/d_{\mathrm{pe}}}},
\end{equation}
where \( i \) is the dimension index, and \( d_{\mathrm{pe}} \) is the total number of positional encoding dimensions.

As an alternative, we adopt the learnable Time Modulation (TM) approach proposed by \citet{cabrera2024atat}, which incorporates temporal information directly into the input features. Instead of using a fixed linear projection, we pass each input vector \( x_{j,b} \) through a MoE layer to obtain an adaptive representation. The resulting vector is then modulated using band-specific, time-dependent Fourier functions:

The resulting vector is then modulated using band-specific, time-dependent Fourier functions:
\begin{equation}
\mathrm{TM}(x_{j,b}, t_{j,b}) = \mathrm{MoE}(x_{j,b}) \odot \gamma^{(1)}_b(t_{j,b}) + \gamma^{(2)}_b(t_{j,b}),
\end{equation}
\vskip -0.1in
where \( \gamma^{(1)}_b \) and \( \gamma^{(2)}_b \) are learnable band-specific Fourier series. The operator \( \odot \) denotes element-wise (Hadamard) multiplication.

\section{Experiments}

\subsection{Data description}
For pretraining and classification, we use data from the first round of the Extended LSST Astronomical Time-series Classification Challenge (ELAsTiCC-1), a large-scale simulation designed to emulate the observational characteristics of the Vera C. Rubin Observatory’s LSST. The dataset contains 1{,}845{,}146 multiband light curves spanning 32 astrophysical classes, including both periodic variables and transient events. Each light curve is observed in six optical bands (\textit{ugrizy}) with realistic cadences, noise levels, and detection limits. Following \citet{cabrera2024atat}, we regroup the classes into 20 categories, discard poor-quality measurements using \texttt{PHOTFLAG}, and extract forced photometry ranging from 30 days before the first alert to the final detection. Additionally, each object is associated with 64 metadata columns describing contextual and observational properties. These include redshift estimates, sky coordinates, host galaxy characteristics, and summary statistics of the light curves. As in previous work, these metadata are incorporated as complementary inputs for classification. The test set contains 1{,}000 objects per class, ensuring balance across categories, while the remaining data are split into five class-stratified folds with an 80/20 training-validation ratio.

To evaluate our approach on an alternative  classification task, we use photometric light curves from the second data release of Pan-STARRS1 (PS1), which offers observations in five optical bands: \textit{g\textsubscript{P1}}, \textit{r\textsubscript{P1}}, \textit{i\textsubscript{P1}}, \textit{z\textsubscript{P1}}, and \textit{y\textsubscript{P1}}. Following the methodology of \citet{becker2025multiband}, we retrieve the PS1 photometry from the Detections table via MAST CasJobs, convert fluxes to AB magnitudes using the standard zero-point of 3631~Jy, and apply quality filters (e.g., \texttt{psfQfPerfect}, \texttt{infoFlags}, \texttt{infoFlags2}, \texttt{infoFlags3}) to ensure clean photometry. We require a minimum of four observations per band. As in \citet{becker2025multiband}, we apply class balancing by limiting the maximum number of objects per class to 10{,}000, mitigating overfitting due to the strong class imbalance, particularly for RR Lyrae stars. Our final dataset includes six variable star classes. The data are split into seven folds, stratified by class, using 70\% for training, 10\% for validation, and 20\% for testing, with the test set kept fixed across all folds. Appendix~\ref{app:clf} shows the classes and number of objects in ELAsTiCC-1 and PS1, grouped as transients, stochastic variables, and periodic variables.

\subsection{Training details}


All model variants are based on an encoder-only transformer architecture with three self-attention blocks. Pretraining is performed using a masked reconstruction objective, following the strategy introduced by \citet{donoso2022astromer}. In each training step, 90\% of the light curves are randomly selected for training. Within each selected light curve, 30\% of the input tokens are masked, 30\% are replaced with random values, and the remaining 40\% are left unchanged. A dropout rate of 0.1 is applied throughout the network. For ELAsTiCC-1 classification, we concatenate the light curve embeddings with tabular embeddings extracted from a tabular transformer, following the ATAT architecture proposed by \citet{cabrera2024atat}. For the PS1 classification task, we apply a linear classifier directly on the light curve embeddings, without incorporating metadata. In both tasks, models are trained using a cross-entropy loss. Training is performed with the Adam optimizer, using a batch size of 256 and a learning rate of $1 \times 10^{-4}$.

\section{Results}

\begin{table}[t]
\caption{Pretraining performance on the ELAsTiCC-1 test set.}
\label{sample-table1}
\begin{center}
\begin{small}
\begin{sc}
\begin{tabular}{lccr}
\toprule
Model   &TE & \( R^2 \) & RMSE \\
\midrule
Multiband-Astromer  &PE & 0.349 & 2.511 \\
\midrule
MoE-Astro   &PE & 0.403 & 2.398 \\
MoE-Astro  &TM & \textbf{0.438} & \textbf{2.327} \\
\bottomrule
\end{tabular}
\end{sc}
\end{small}
\end{center}
\vskip -0.15in
\end{table}

\begin{table}[b]
\small
\caption{Classification performance on the ELAsTiCC-1 test set using both light curve and metadata information.}
\label{sample-table2}
\vskip -0.8in
\begin{center}
\begin{small}
\begin{sc}
\begin{tabular}{lcccr}
\toprule
Model   & Pretrained   & F1-score  \\
\midrule
Multiband-  & Yes  & 0.727 $\pm$ 0.034  \\
Astromer  & No  &  0.786 $\pm$ 0.013 \\
\midrule
ATAT  & No  &  0.826 $\pm$ 0.005   \\
\midrule
MoE-Astro (PE)  & Yes  &    0.822 $\pm$ 0.008 \\
MoE-Astro (TM)  & Yes  &    0.832 $\pm$ 0.015\\
MoE-Astro (TM)   & No &   \textbf{0.860} $\pm$ \textbf{0.003} \\
\bottomrule
\end{tabular}
\end{sc}
\end{small}
\end{center}
\end{table}

\begin{table}[t]
\small
\caption{Classification performance on the Pan-STARRS1 test set.}
\label{sample-table3}
\begin{center}
\begin{small}
\begin{sc}
\begin{tabular}{lcc}
\toprule
Model & Pretrained & F1-score \\
\midrule
MoE-Astro (TM) & No & 0.373 $\pm$ 0.007 \\
MoE-Astro (TM) & Yes & \textbf{0.542} $\pm$ \textbf{0.023} \\
\bottomrule
\end{tabular}
\end{sc}
\end{small}
\end{center}
\vskip -0.3in
\end{table}

Table~\ref{sample-table1} reports the mean and standard deviation of pretraining scores for our model configurations on the ELAsTiCC-1 test set. As a baseline, we consider a multiband extension of Astromer, in which brightness vectors are ordered by observation time and projected via a shared linear layer. Comparing this baseline with our proposed Astro-MoE architecture, we observe improvements when incorporating MoE modules for both the brightness encoder and the FFN within the attention blocks. Specifically, this configuration achieves an \( R^2 \) of 0.403 and an RMSE of 2.398. When replacing the PE with the TM encoder, performance further improves by approximately 3\%, reaching an \( R^2 \) of 0.438 and reducing RMSE to 2.327. Overall, these results indicate that the combination of sparse MoE-based representations and temporally adaptive encoding may enhance the expressiveness and accuracy of light curve models. It is also important to note that ELAsTiCC-1 is a complex dataset, featuring a maximum sequence length of 65 per band and comprising a wide diversity of astronomical object classes.

After pretraining, we evaluate the models in a classification setting. Table~\ref{sample-table2} reports the mean and standard deviation of macro F$_1$-scores on the ELAsTiCC-1 dataset using a multimodal approach that integrates both light curve and metadata features. The pretrained Multiband-Astromer baseline underperforms compared to its non-pretrained version (F$_1 = 0.727 \pm 0.034$ vs. F$_1 = 0.786 \pm 0.013$), suggesting that pretraining on the same dataset may result in early convergence and reduced adaptability during downstream fine-tuning. We also include ATAT~\citep{cabrera2024atat}, a non-pretrained transformer tailored for multimodal inputs, which achieves a competitive score of (F$_1 = 0.826 \pm 0.005 $) under identical evaluation conditions. Our MoE-Astro model, which incorporates sparse expert routing in place of standard transformer components, yields consistent performance improvements. Specifically, the pretrained MoE-Astro with positional encoding (PE) reaches (F$_1 = 0.822 \pm 0.008$), while its variant with time modulation improves to (F$_1 = 0.832 \pm 0.015 $). Notably, when trained from scratch, MoE-Astro (TM) achieves the highest performance with (F$_1 = 0.860 \pm 0.003 $). These results suggest that MoE-based architectures can provide enhanced capacity allocation and generalization (see Appendix~\ref{app:cm}). While the benefits of pretraining may depend on task similarity and dataset diversity, our approach currently achieves state-of-the-art performance on ELAsTiCC-1.

To assess the transferability of the best pretrained model (Astro-MoE with TM), we evaluate it on the challenging and imbalanced PS1 dataset. As shown in Table~\ref{sample-table3}, pretraining improves performance from an F\(_1\) score of 0.373 to 0.542, suggesting that Astro-MoE is capable of generalizing to new domains. This preliminary result highlights its potential for cross-survey applications and motivates further exploration across diverse real-world datasets.

\section{Conclusion}

We have empirically found evidence that sparse MoE models offer clear benefits for multiband astronomical time series analysis, with advantages observable even at initial scales. Their ability to allocate capacity dynamically, combined with efficient computation, makes them well-suited for large-scale pretraining on the heterogeneous and growing datasets of time-domain astronomy. Beyond improving performance, these architectures also offer practical benefits: by activating only a subset of experts per input, they reduce the computational cost during inference. This property makes them particularly attractive for real-time applications, such as transient classification in astronomical alert streams. Overall, Astro-MoE represents a promising direction for developing scalable and adaptive models to support the next generation of astronomical discovery.

\section*{Acknowledgments}
The authors acknowledge support from the National Agency for Research and Development (ANID) grants: FONDECYT regular 1231877 (GCV, DMC); Millennium Science Initiative Program ICN12\_009 (GCV). The computations in this paper were run on the FASRC Cannon cluster supported by the FAS Division of Science Research Computing Group at Harvard University. 

\bibliography{example_paper}
\bibliographystyle{icml2025}

\newpage
\appendix
\onecolumn

\section{Classification scheme}
\label{app:clf}
Table~\ref{table:elasitcc_ps1_classes} provides an overview of the classification taxonomy used in our work, summarizing the number of objects per class across two datasets: ELAsTiCC-1 and Pan-STARRS1 (PS1). The ELAsTiCC-1 taxonomy is structured into three primary variability types (transient, stochastic, and periodic) encompassing a wide range of astrophysical phenomena including supernovae subtypes (e.g., Ia, II, Iax), cataclysmic variables (e.g., Dwarf Novae), and pulsating stars (e.g., Delta Scuti, RR Lyrae). The PS1 dataset, by contrast, focuses exclusively on periodic variables, reflecting its strengths in long-term monitoring of the sky. The pronounced class imbalance poses challenges for training robust machine learning classifiers and underscores the importance of methods capable of handling skewed data distributions.

\begin{table}[h]
\caption{Number of objects per class in ELAsTiCC-1 and PS1, grouped by variability type. Each group lists the included classes and the number of objects in parentheses.}
\label{table:elasitcc_ps1_classes}
\vskip 0.1in
\begin{center}
\renewcommand{\arraystretch}{1.3}
\begin{footnotesize}
\begin{tabular}{|p{2cm}|p{5.5cm}|}
\toprule
\textbf{Group} & \textbf{Classes (number of objects)} \\
\midrule
\multirow{6}{*}{\shortstack[l]{ELAsTiCC-1\\Transient (12)}} & CART (15{,}719), Iax (53{,}727) \\
& 91bg (53{,}414), Ia (211{,}892) \\
& Ib/c (310{,}328), II (445{,}419) \\
& SN-like/Other (103{,}683), SLSN (105{,}238) \\
& PISN (105{,}446), TDE (103{,}067) \\
& ILOT (14{,}253), KN (8{,}122) \\
\midrule
\multirow{2}{*}{\shortstack[l]{ELAsTiCC-1\\Stochastic (4)}} & M-dwarf Flare (2,640), uLens (27{,}263) \\
& Dwarf Novae (12{,}385), AGN (99{,}461) \\
\midrule
\multirow{2}{*}{\shortstack[l]{ELAsTiCC-1\\Periodic (4)}} & Delta Scuti (29{,}840), RR Lyrae (21{,}100) \\
& Cepheid (25{,}371), EB (96{,}778) \\
\midrule
\multirow{3}{*}{\shortstack[l]{Pan-STARRS1\\Periodic (6)}} & RRab (10{,}000), RRc (10{,}000) \\
& RRd (266), MIRA\_SR (3{,}937) \\
& DSCT\_SXPHE (1{,}906), T2CEP (189) \\
\bottomrule
\end{tabular}
\end{footnotesize}
\end{center}
\vskip -0.15in
\end{table}

\section{Astro-MoE confusion matrices}
\label{app:cm}
Figures 2, 3, and 4 present detailed visualizations of our proposed sparse MoE model’s classification performance across various astronomical object types. To ensure robust and interpretable insights, we analyze confusion matrices aggregated over multiple evaluation runs. Each matrix is row-normalized so that each row sums to 100\%, reflecting the distribution of predicted classes for each true class label. We report the median values across runs to provide a stable central estimate, which is less affected by outliers compared to the mean. To capture variability, we also compute the 25th and 95th percentiles, offering a clear view of the range in classification performance.

Each cell in the visualization conveys three metrics: the median prediction percentage (center), the 95th percentile (top right), and the 25th percentile (bottom right). Diagonal cells represent correct classifications, while off-diagonal cells indicate misclassifications. The color intensity encodes the prediction percentage, with darker green denoting higher accuracy. For visual clarity, cells with values below 0.05\% are left blank.

\begin{figure}
\begin{center}
\label{fig:fig1}
\includegraphics[width=18cm]{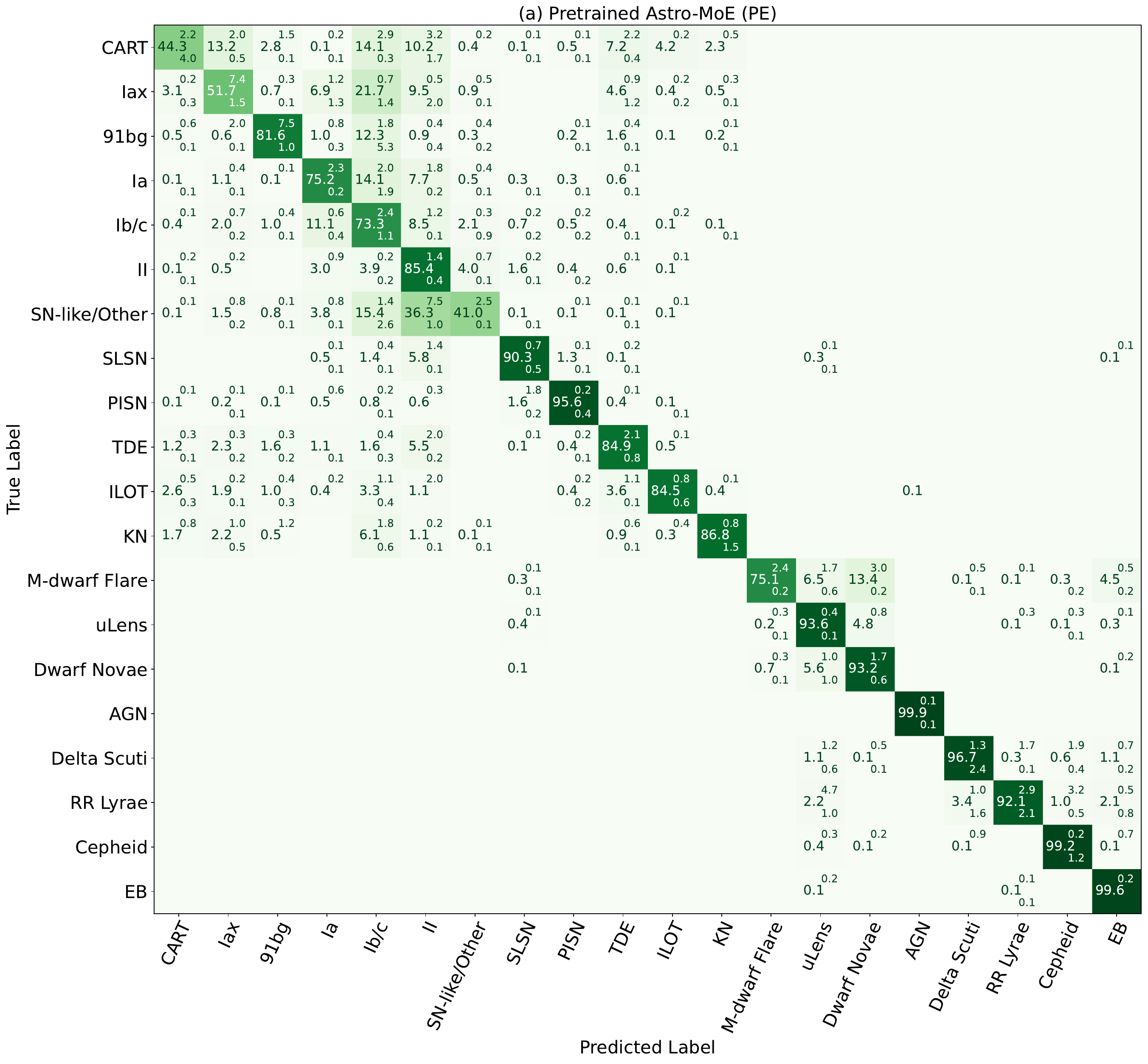}
\caption{Pretrained Astro-MoE (PE) confusion matrix.}
\end{center}
\end{figure}

\begin{figure}
\begin{center}

\includegraphics[width=18cm]{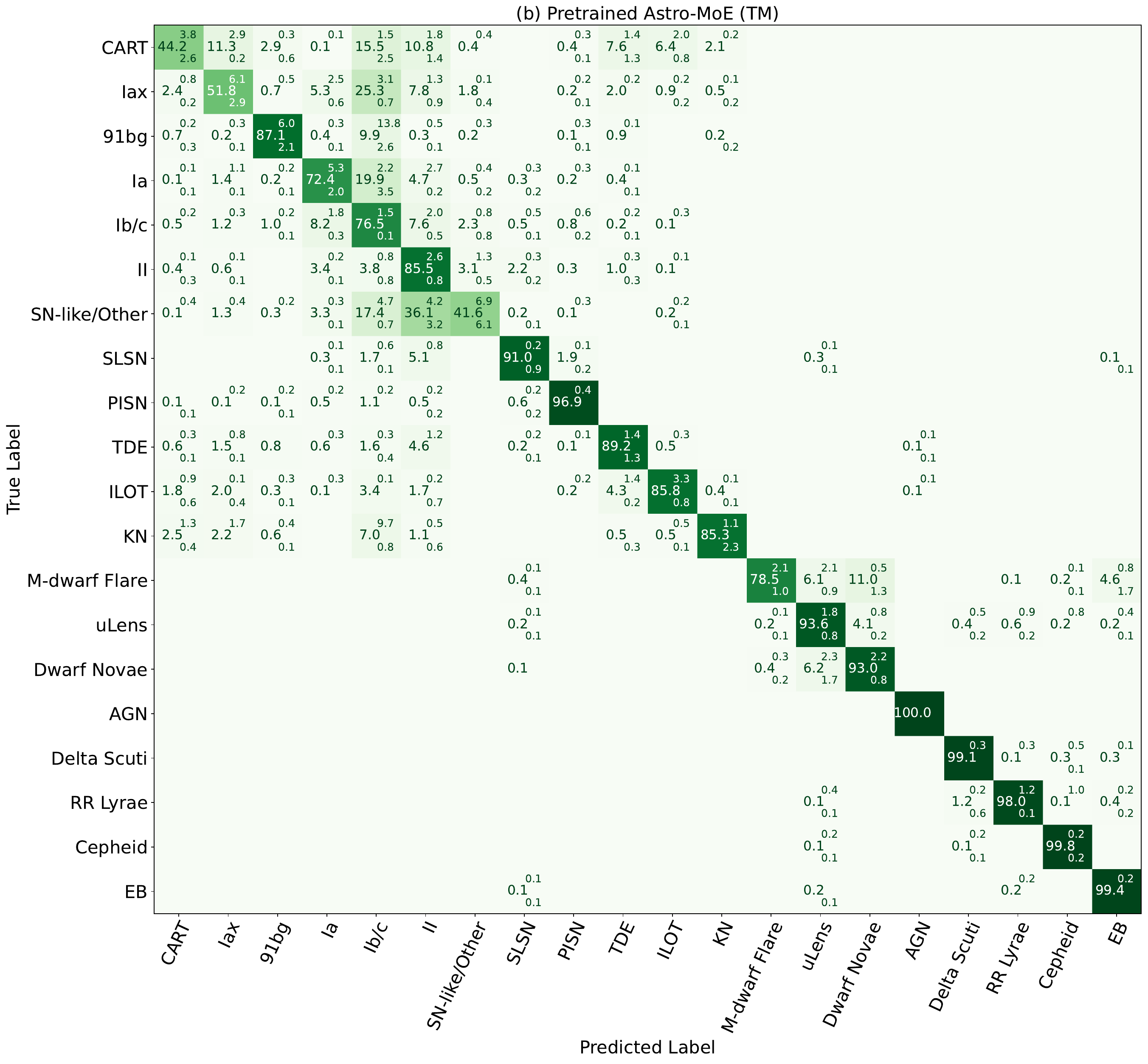}
\caption{Pretrained Astro-MoE (TM) confusion matrix.}
\end{center}
\label{fig:fig2}
\end{figure}

\begin{figure}
\begin{center}
\includegraphics[width=18cm]{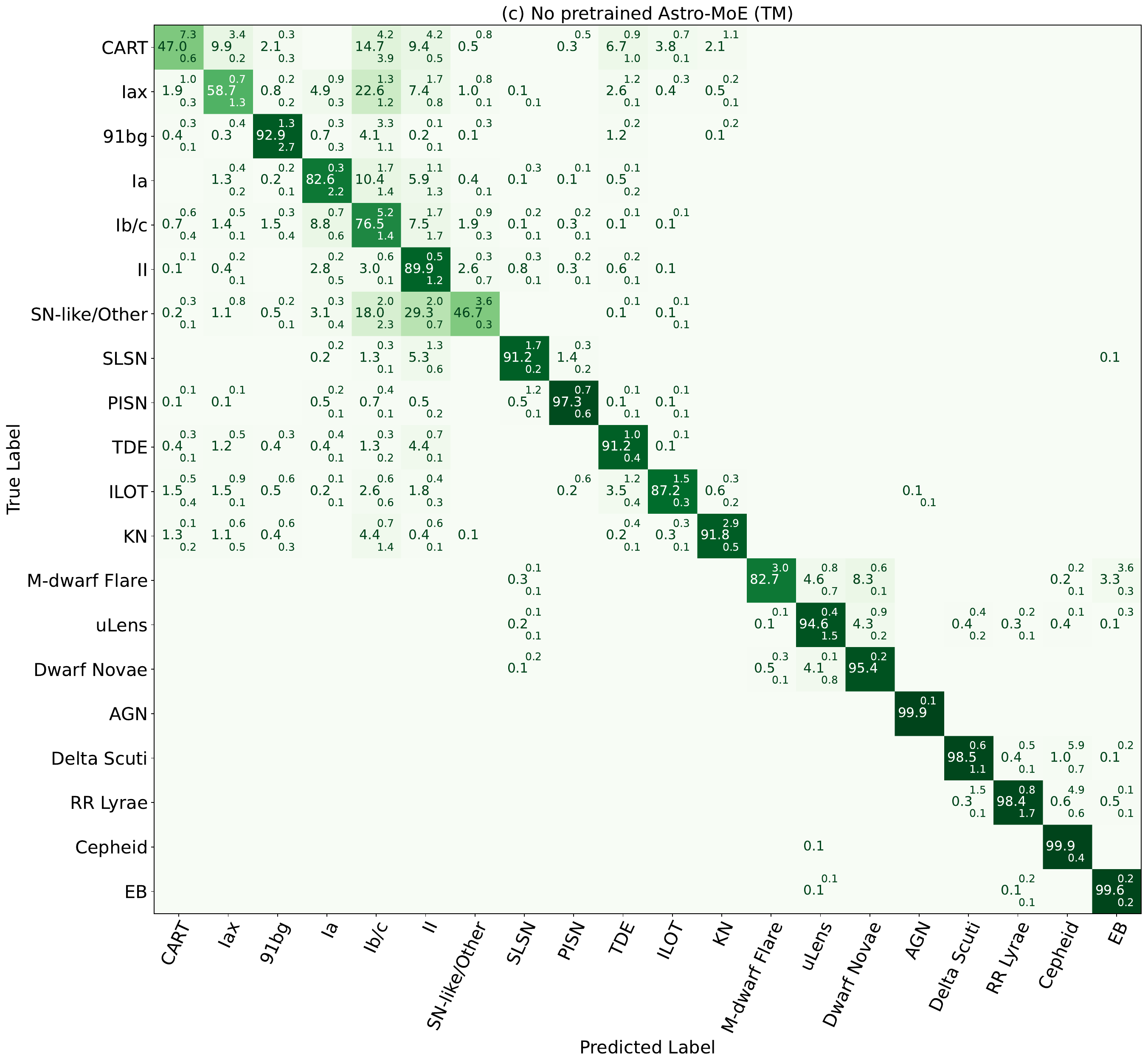}
\caption{Non-pretrained Astro-MoE (TM) confusion matrix.}
\end{center}
\label{fig:mat3}
\end{figure}

\section{Parameter count comparison}
We introduce a Mixture-of-Experts variant of our model to enable dynamic routing through specialized feedforward layers. This architecture substantially increases model capacity, with approximately 3× more total parameters (4.9M vs. 1.5M). However, thanks to its sparse activation mechanism, the GPU inference time per batch increases only modestly, from 5.3ms to 7.1ms.

Critically, the MoE model activates only a subset of experts per forward pass, specifically, 2 out of 8 experts in the feedforward network and 6 in the positional embedding layer. This sparsity allows the model to retain high expressiveness without incurring the full computational cost of using all parameters. Moreover, this design opens up avenues for further optimization: reducing the number of active experts, for instance, could further lower inference time without necessarily compromising performance.

This tradeoff highlights a core advantage of MoE architectures; their ability to scale capacity without a linear increase in computation or latency. In the context of ELAsTiCC, where classification involves a wide range of astrophysical phenomena and complex temporal dynamics, such additional representational power is beneficial. Overall, our results suggest that the increased capacity of the MoE model justifies the minor overhead, providing a scalable and efficient solution for modeling heterogeneous multiband time series data.

\label{sec:pcc}
\begin{table}[t]
\caption{Comparison of model capacity and inference efficiency across variants for the ELAsTiCC classification task.}
\begin{center}
\begin{small}
\begin{sc}
\begin{tabular}{lccr}
\toprule
Model   & TE & Params & Inference time (ms) \\
\midrule
Multiband-Astromer  & PE & 1.5M  & 5.3 ms \\
\midrule
MoE-Astro   &PE &  4.6M & 6.8 ms \\
MoE-Astro  &TM & 4.9M  & 7.1 ms  \\
\bottomrule
\end{tabular}
\end{sc}
\end{small}
\end{center}
\vskip -0.15in
\end{table}

\end{document}